\documentclass[aps,prd,showpacs]{revtex4}

\usepackage{graphicx}
\usepackage{undertilde}

\newcommand{\xPert}{{\em \lowercase{x}P\lowercase{ert}}}
\newcommand{\pert}[2]{{}^{\mbox{\,\tiny $\{#1\}\!$}}{#2}}
\newcommand{\h}[1]{{}^{\mbox{\,\tiny $\{#1\}\!$}}h}
\newcommand{\binomial}[2]{
    \left( \begin{array}{c} {#1} \\ {#2} \end{array} \right)
}
\newcommand{\mathIn}[2]{
    \begin{quotation} {\em In[#1] :=\ } {\tt #2} \end{quotation}
}
\newcommand{\mathOut}[2]{
    \begin{quotation} {\em Out[#1] :=\ } #2 \end{quotation}
}
\newcommand{\mathCode}[1]{
    \begin{quotation} \qquad\qquad {\tt #1 } \end{quotation} 
}
\newcommand{\mathComment}[1]{
    \begin{quotation} \qquad\qquad #1 \end{quotation} 
}

\begin{document}

\title{\xPert\/: Computer algebra for metric perturbation theory}


\author{David Brizuela$^1$}
\author{Jos\'e M. Mart\'{\i}n-Garc\'{\i}a$^{1,2}$}
\author{Guillermo A. Mena Marug\'an$^1$}
\affiliation{${^1}$Instituto de Estructura de la Materia, CSIC,
Serrano 121-123, Madrid 28006, Spain}
\affiliation{${^2}$Matematiska institutionen, Link\"opings universitet,
Link\"oping, Sweden S-581 83}

\begin{abstract}
We present the tensor computer algebra package \xPert\/ for fast
construction and manipulation of the equations of metric perturbation
theory, around arbitrary backgrounds. It is based on the combination
of explicit combinatorial formulas for the $n$-th order perturbation
of curvature tensors and their gauge changes, and the use of highly
efficient techniques of index canonicalization, provided by the
underlying tensor system {\em xAct}, for {\em Mathematica}.
We give examples of use and show the efficiency of the system with
timings plots: it is possible to handle orders $n=4$ or $n=5$ within
seconds, or reach $n=10$ with timings below 1 hour.
\end{abstract}

\maketitle


\section{Introduction}

{\em Linear} perturbation theory is one of the most successful techniques
to extract information from the equations of general relativity.
It allows us to study the evolution of small deviations with respect to
a known background solution of the equations, showing the stability
properties of that background or how it interacts with weak surrounding
fields. Linearization is, however, the main limitation of the approach,
because we loose the intrinsic nonlinear character of general relativity,
and hence it is not possible to model the self-interaction or mutual
interaction among the linear modes, and other interesting phenomena.
Obviously, this could be cured by adding a second perturbative order to
the analysis, but the large size of the resulting perturbative equations
has generally prevented practical computations until very recently.

Already in 1980, Cunningham, Price and Moncrief studied and simulated
second-order perturbations of collapse of a ball of dust 
(Oppenheimer-Snyder collapse), including the matching to perturbations of an
exterior vacuum \cite{CPM80}. Since then, other backgrounds have been
perturbed at second order. The oscillation modes of a rotating fluid
star have been modelled at second order, introducing rotation at linear
level \cite{CF91, Koj92}.
Teukolsky's formalism \cite{Teu72} for linear perturbations of a
Kerr black hole was generalized to second order in \cite{CaLo99}.
The gravitational radiation emitted in a close limit head-on collision
of two black holes has also been studied at second order, describing
the final hole as a first-order quadrupolar perturbation of the
Schwarzschild spacetime \cite{PrPu94, GNP96}. Recently, also perturbing
Schwarzschild, second-order quasinormal modes have been
defined \cite{IoNa07, NaIo07}. Second-order perturbations
have been considered in several cosmological scenarios, often
taking Friedmann-Robertson-Walker as background spacetime;
see for example \cite{MPS94, SSC94, BMT02} and
references therein. We can also find high-order perturbations in
various areas of field theory, like the problem of
renormalization of a Lagrangian \cite{BFT}, though in this context the
formalism used is slightly different. We shall later see how the general
scheme can be adapted to include this as a particular case.

In references \cite{BMM06, BMM07} we have constructed the arbitrary-order
generalization of the Gerlach and Sengupta formalism for gauge-invariant
and covariant first-order perturbations of any spherical spacetime
\cite{GeSe79}, for any matter model. This is the most convenient framework to study
nearly-spherical astrophysical scenarios, like collapse of slowly-rotating
stars, and some types of supernova explosions. The large size of the
equations involved forced us to develop specialized tools of tensor
computer algebra, which were summarized in an appendix of \cite{BMM06}.
We have now improved these tools, extending their domain of applicability,
increasing their efficiency, simplifying their use and fully documenting
them.

We present version 1.0 of the package \xPert\/ for high-order metric
perturbation theory. It is based on the combination of adapted
mathematical algorithms and powerful techniques of tensor computer
algebra. On the one hand, we use explicit precomputed formulas for the
general terms of the perturbative expansions of the most important
curvature tensors in differential geometry, avoiding slow recursive
computations. On the other hand, \xPert\/ uses very efficient
canonicalization algorithms for tensor computer algebra in
{\em Mathematica}. This allows us to perform computations up to
perturbative order 10 in several minutes with small computer resources,
as we will see in some examples.

The article is organized as follows. Section \ref{theory} reviews
the basic concepts of high-order metric perturbation theory and
presents closed formulas to obtain the $n$-th perturbation of
the most relevant geometric tensors. 
Section \ref{code} explains how these formulas have been implemented
in {\em Mathematica} and lists the main commands of the package \xPert.
Finally, we show the high efficiency of the system through some timing
examples in section \ref{examples}. Section \ref{conclusions}
contains our conclusions.

\section{High-order metric perturbation theory}\label{theory}

\subsection{Basic concepts}

Let us assume a family of $N$-dimensional manifolds
$\overline{\cal M}(\varepsilon)$ equipped with a metric field
$\overline{g}(\varepsilon)$ depending on a dimensionless parameter
$\varepsilon$. This dependence will be taken to be smooth,
so that all derivatives with respect to $\varepsilon$ are well defined.
Members with parameter $\varepsilon=0$ will be referred to as
{\em background} members and will be denoted without bar,
that is ${\cal M}\equiv \overline{\cal M}(0), g\equiv \overline g(0)$.
The Levi-Civita connection of $g$ will be called $\nabla$, and
will be denoted with a semicolon in indexed expressions.

The goal in perturbation theory is comparing a tensor field
$\overline{T}(\varepsilon)$ with its background counterpart $T$. The former is
defined on the manifold $\overline{\cal M}(\varepsilon)$ and the latter on
the background manifold ${\cal M}$ and therefore we have to introduce a
family of point-to-point identification mappings (or {\em gauge}) $\phi_\varepsilon$
between them to compare. The pull-back $\phi_\varepsilon^*\overline{T}(\varepsilon)$ to
the background can be now expanded as
\begin{equation}
\phi_\varepsilon^*\overline{T}(\varepsilon)\equiv T +
             \sum_{n=1}^{\infty}\frac{\varepsilon^n}{n!}\Delta_\phi^n[T],
\end{equation}
with all terms in the equation being defined on the background manifold.
We have introduced the perturbative operator $\Delta_\phi$ in the gauge
$\phi_\varepsilon$, which is actually a derivative:
\begin{equation}
\Delta_\phi^n[T]\equiv \left.
\frac{d^n \, \phi_\varepsilon^*\overline{T}(\varepsilon)}{d\varepsilon^n}
\right|_{\varepsilon=0}.
\end{equation}
From now on we shall omit the explicit dependence of perturbations on
the chosen gauge $\phi_\varepsilon$, to alleviate the notation. $\varepsilon$-dependent
(barred) objects will be also assumed to be pulled-back to the background
manifold. The theory and formulas for perturbations of changes of gauge
have been fully developed for tensors by Bruni and collaborators
\cite{BMM97}, and extended in \cite{Pit07} to perturbations of
distribution functions in phase space.

An important consequence of the derivative character of $\Delta$ is that
it obeys the Leibnitz rule, whose $n$-th order generalization on the
product of $m$ tensors is
\begin{equation}\label{product} 
\Delta^n[T_1\dots T_m]=\sum_{\{k_i\}} \frac{n!}{k_1!\dots k_m!}\Delta^{k_1}[T_1]\dots\Delta^{k_m}[T_m],
\end{equation}
where the sum extends to all sorted partitions of $n$ in $m$ nonnegative
integers (that is, including zero) obeying $k_1+...+k_m=n$.
It is possible to generalize this formula to the case
$\Delta^n[ F(S_1, ..., S_m) ]$ for an arbitrary function $F$ of $m$
scalar arguments $S_1,..., S_m$ using the
Fa\`a di Bruno formula \cite{AbSt} (the $n$-th order chain rule),
generalized to multiple arguments.

We want to do metric perturbation theory, meaning that our fundamental
object to be perturbed is the metric field $g_{ab}$ (abstract indices
in any of the tangent spaces of the $\overline{\cal M}$ manifolds will be
lowercase Latin indices). It is convenient to
name the $n$-th order perturbation of the metric as
$\pert{n}{h}_{ab}\equiv \Delta^n[g_{ab}]$, with a left-superindex
denoting the perturbative order. In this notation, the perturbed metric
is given by
\begin{equation} \label{metricexpansion}
\overline g_{ab}(\varepsilon)=g_{ab} + \sum_{n=1}^\infty\frac{\varepsilon^n}{n!} \pert{n}{h}_{ab}.
\end{equation}
With nonvanishing perturbations of the background metric, it is essential
to carefully keep track of index positions. For example we have
$\Delta[v^a]\neq g^{ab}\Delta[v_b]$ for a tensor field $v^a$. The notation
$\Delta v^a$ could be ambiguous and therefore we use square brackets in
our notation for perturbations to avoid problems. Should we find the
perturbation of a tensor with inconvenient index positions, we shall
introduce metric factors to correct those positions and then use the
Leibnitz rule (\ref{product}).

An important variation of this general perturbative formalism
is frequently used in quantum field theory in curved backgrounds.
It is usually called `background field method' and decomposes the
full metric $\overline{g}$ as $\overline{g}_{ab} = g_{ab} + h_{ab}$ \cite{BFT}.
Other objects depending nonlinearly on $\overline{g}$ are later truncated at
a given order in powers of $h$. This can be considered as a particular
case of the general formalism used here, in which all but the first
perturbations of the covariant metric are defined to be zero:
$\h{n}_{ab}=0$ for $n\geq 2$. All formulas in this article can be
translated to background field method using such a simple restriction.

\subsection{Perturbations of curvature tensors}

The first step in any perturbative computation in general relativity is
expressing the perturbation of one or several curvature tensors in terms
of perturbations of the metric. This is a straightforward but very long
computation for high orders, and hence requires the use of computer
algebra tools like \xPert. There are essentially two ways to
perform such a computation: either recursively, computing the $n$-th
order perturbation from the previous order, or using explicit
formulas giving the result directly. \xPert\/ implements such
explicit formulas for the most important tensors in general relativity,
as shown in this and the next subsections. Some of these formulas have
been already presented in \cite{BMM06}.

The general term of the expansion of the inverse metric $\overline{g}^{ab}$
as a power series in $\varepsilon$ can be obtained as follows.
Let us define $H_{ab}\equiv\overline{g}_{ab}-g_{ab}$ and temporarily drop
the indices. Then, the inverse of the metric is given by
\begin{equation}
\overline g^{-1}=(g+H)^{-1},
\end{equation}
which can be expanded around $H=0$ as
\begin{equation}
\overline g^{-1}=g^{-1}\sum_{m=0}^{\infty}(-1)^m(g^{-1}H)^m.
\end{equation}
Reintroducing indices and expanding $H$ itself using (\ref{metricexpansion})
we obtain the general term
\begin{equation}\label{invmetricexpansion}
\Delta^n[g^{ab}] =
\sum_{(k_i)} (-1)^m\frac{n!}{k_1!\,...\,k_m!} \h{k_m}^{ae_m}\;\h{k_{m-1}}_{e_m}{}^{e_{m-1}}\;...\;
\h{k_2}_{e_3}{}^{e_2}\;\h{k_1}_{e_2}{}^b,
\end{equation}
where the sum $\sum_{(k_i)}$ runs over the $2^{n-1}$ sorted partitions of
$n$ in $m\le n$ positive integers (not including zero this time) such that
$k_1+...+k_m=n$.

With the general term of the expansion of the metric and its inverse at
hand, we can obtain the general term for the perturbation of the
Christoffel symbols starting from its definition in terms of the metric:
\begin{equation} \label{Christoffelexpansion}
\Delta^n[\Gamma^a{}_{bc}] =
\sum_{(k_i)} (-1)^{m+1}\frac{n!}{k_1!\,...\,k_m!}
\h{k_m}^{ae_m}\;
\h{k_{m-1}}_{e_m}{}^{e_{m-1}}\;...\;
\h{k_{2}}_{e_3}{}^{e_2}\;\h{k_1}_{e_2bc},
\end{equation}
where the three-indices tensor perturbation is given by
\begin{equation}
\h{n}_{abc} \equiv \frac{1}{2}\left( \h{n}_{ab ;c}
+ \h{n}_{ac ;b} - \h{n}_{bc ;a}\right),
\end{equation}
which is symmetric in its last two indices. For $n=1$ formula 
(\ref{Christoffelexpansion}) must be read
as $\Delta[\Gamma^a{}_{bc}]=\h{1}^{a}{}_{bc}$. Note that the perturbation
of the Christoffel symbols is constructed from the metric perturbations
$\pert{n}{h}_{ab}$ and their first covariant derivative at all orders.

The perturbation of the Riemann tensor of an arbitrary connection (not
necessarily a metric connection) is a simple application of the
Leibnitz rule (\ref{product}):
\begin{equation}\label{pertRiemann}
\Delta^n[R_{abc}{}^d] =
\nabla_b\left(\Delta^n[\Gamma^d{}_{ac}]\right) -
\sum_{k=1}^{n-1} \binomial{n}{k}
\Delta^k[\Gamma^e{}_{bc}]
\Delta^{n-k}[\Gamma^d{}_{ea}] -
(a\leftrightarrow b).
\end{equation}
For a metric-compatible connection we can use (\ref{Christoffelexpansion}),
getting
\begin{eqnarray} \label{pertRiemann2}
\Delta^n[R_{abc}{}^d] &=&
\sum_{(k_i)} (-1)^m \frac{n!}{k_1!...k_m!} \Big[
\h{k_m}^{de_m}...\h{k_2}_{e_3}{}^{e_2}
\h{k_1}_{e_2cb;a} \\
 && \qquad + \sum_{s=2}^{m}
\h{k_m}^{de_m}...
\h{k_{s+1}}_{e_{s+2}}{}^{e_{s+1}}
\h{k_s}_{e_se_{s+1}a}
\h{k_{s-1}}^{e_se_{s-1}}...
\h{k_2}_{e_3}{}^{e_2}
\h{k_1}_{e_2bc} \Big]
- (a\leftrightarrow b). \nonumber
\end{eqnarray}

Other relevant curvature tensors are given in terms of Riemann, and
therefore their perturbations can be expressed in terms of the latter
formula. For instance, perturbations of Ricci are given by
\begin{equation}\label{pertRicci}
\Delta^n[R_{ab}] = \Delta^n[R_{acb}{}^c].
\end{equation}
The Ricci scalar $R\equiv g^{ab}R_{ab}$ can be perturbed using the
generalized Leibnitz rule (\ref{product}):
\begin{equation}\label{pertRicciScalar}
\Delta^n[R] = \sum_{k=0}^n \binomial{n}{k}
\Delta^k[g^{ab}]\Delta^{n-k}[R_{ab}] ,
\end{equation}
and similarly for the Einstein tensor:
\begin{equation}\label{pertEinstein}
\Delta^n[G_{ab}]=\Delta^n[R_{ab}]
-\frac{1}{2}\sum_{k=0}^{n}\sum_{j=0}^{k}
\frac{n!}{k!\,j!\,(n-j-k)!}
\h{j}_{ab}\,\Delta^k[g^{cd}]\,\Delta^{n-j-k}[R_{cd}].
\end{equation}
We have not tried to further simplify these two last formulas, as they
are already quite efficient and give directly the $n$-th term of the
respective perturbative expansions.

\subsection{Perturbation of the metric determinant}
In Hamiltonian field theory we frequently find the perturbation of the
determinant of the metric. This is a basis-dependent concept, in the
sense that what we are computing is the determinant of the components of
the metric in a given basis and the result depends on the basis we have
chosen. Under a change of basis the determinant changes with a squared
Jacobian and it is, hence, a {\em density} of {\em weight} +2.

The determinant of the metric $g_{ab}$ can be defined as
\begin{equation}
\det(g_{ab})\equiv \frac{1}{N!}\,\widetilde\eta^{a_1\dots a_N}\,\widetilde\eta^{b_1\dots b_N}\,g_{a_1b_1}\dots g_{a_Nb_N},
\end{equation}
using the upper antisymmetric density $\widetilde\eta^{a_1\dots a_N}$
(the overtilde denotes weight +1), whose components
in the chosen basis are $+1$, $-1$ or $0$. This object, as well as its
lower counterpart $\utilde\eta_{a_1\dots a_N}$ of weight $-1$, stays invariant under
the $\Delta$ perturbation. Therefore, the Leibnitz rule (\ref{product}) implies
\begin{equation}\label{pertDet}
\Delta^n[\det(g_{ab})]=\frac{1}{N!}\,
\widetilde\eta^{a_1\dots a_N}\,\widetilde\eta^{b_1\dots b_N}
\sum_{\{k_i\}} \frac{n!}{k_1!\dots k_N!}\pert{k_1}{h}_{a_1b_1}\dots\pert{k_N}{h}_{a_Nb_N}.
\end{equation}
Finally, this can be simplified using the well known relation
\begin{equation}
\widetilde\eta^{a_1\dots a_N}\,\widetilde\eta^{b_1\dots b_N}
= \det(g_{ab})
\left|
\begin{array}{ccc}
g^{a_1b_1} & \cdots & g^{a_1b_N} \\
\vdots & & \vdots \\
g^{a_Nb_1} & \cdots & g^{a_Nb_N} 
\end{array}
\right| .
\end{equation}
We conclude that the $n$-th order perturbation of the determinant of the
metric is always the product of the determinant itself times a scalar
formed by contraction of metric perturbations.
It is interesting to note that such a scalar factor contains the product
of at most $N$ metric perturbations, and not $n$, as we might have
anticipated by inspection of the formulas in the previous section.
This is actually the only place in this article in which the dimension
of the manifold being perturbed plays a role.

With this formula at hand it is straightforward to give the
$n$-th order perturbation of the volume form
$\epsilon_{a_1\dots a_N}\equiv
|\det(g)|^{1/2}\, \utilde\eta_{a_1\dots a_N}$.

\subsection{Perturbations of derivatives}
There are several types of derivatives which may appear in a perturbative
computation. We shall study here three of those types: partial,
covariant and Lie derivatives.

Partial derivatives are associated to coordinate systems and hence do not
change under perturbations. Therefore, by construction, they commute with
the $\Delta$ operator: $\Delta^n[T_{,a}]=\Delta^n[T]_{,a}$ for any
tensor field $T$ of any rank.

General covariant derivatives can be perturbed. For instance, the
Levi-Civita connection of a metric will change when its metric is
perturbed. The question arises then about what is the perturbation of
the covariant derivative of a tensor. Transforming to partial derivatives
and Christoffel symbols, perturbing and coming back to covariant
derivatives, we get, for an arbitrary tensor density $T$ of weight $\omega$,
\begin{eqnarray}\label{pertCD}
\Delta^n[T^{a_1\dots a_m}{}_{b_1\dots b_k;c}]&=&\Delta^n[T^{a_1\dots a_m}{}_{b_1\dots b_k}]_{;c} \\ \nonumber
&+&\sum_{i=1}^m \left\{\Delta^n[\Gamma^{a_i}{}_{cd} T^{a_1\dots d\dots a_m}{}_{b_1\dots b_k}]
-\Gamma^{a_i}{}_{cd} \Delta^n[T^{a_1\dots d\dots a_m}{}_{b_1\dots b_k}]\right\}
\\\nonumber
&-&\sum_{j=1}^k\left\{ \Delta^n[\Gamma^{d}{}_{cb_j} T^{a_1\dots a_m}{}_{b_1\dots d\dots b_k}]
-\Gamma^{d}{}_{cb_j} \Delta^n[T^{a_1\dots a_m}{}_{b_1\dots d\dots b_k}]\right\}
\\\nonumber
&-&\omega\;\left\{ \Delta^n[\Gamma^{d}{}_{cd} T^{a_1\dots a_m}{}_{b_1\dots b_k}]
-\Gamma^{d}{}_{cd} \Delta^n[T^{a_1\dots a_m}{}_{b_1\dots b_k}]\right\},
\end{eqnarray}
which can be rewritten applying the Leibnitz rule (\ref{product}) as
\begin{eqnarray}\label{covd}
\Delta^n[T^{a_1\dots a_m}{}_{b_1\dots b_k;c}]&=&
\Delta^n[T^{a_1\dots a_m}{}_{b_1\dots b_k}]_{;c} \\
&+&\sum_{l=1}^n \binomial{n}{l}
\Big\{ 
\sum_{i=1}^m \Delta^l[\Gamma^{a_i}{}_{cd}]\Delta^{n-l}[T^{a_1\dots d\dots a_m}{}_{b_1\dots b_k}]
- \sum_{j=1}^k \Delta^l[\Gamma^{d}{}_{cb_j}]\Delta^{n-l}[ T^{a_1\dots a_m}{}_{b_1\dots d\dots b_k}]
\nonumber \\ \nonumber
&& \qquad\qquad\qquad - \omega\; \Delta^l[\Gamma^{d}{}_{cd}]\Delta^{n-l}[ T^{a_1\dots a_m}{}_{b_1\dots b_k}]
\Big\}.
\end{eqnarray}

Finally, the perturbation formula for the Lie derivative along the vector
field $v$ of a tensor $T$ of any rank can be computed by intermediate
transformation to partial or covariant derivatives:
\begin{equation}\label{pertLieD}
\Delta^n[{\cal L}_{v}T] =
\sum_{k=0}^{n} \binomial{n}{k}
{\cal L}_{\Delta^k[v]}\Delta^{n-k}[T].
\end{equation}
This expression bears an obvious similarity with the Leibnitz rule,
reflecting the fact that both $v$ and $T$ are being perturbed but not
the Lie structure itself, which is directly given by the differential
structure of the manifold and thus remains unperturbed.

\section{The \xPert\/ package}\label{code}

The general terms for the perturbation expansions in the previous section
are combinatorial in nature and hence contain a number of terms which
grows exponentially with the perturbative order $n$. Only through the use
of a specialized tensor computer algebra system it can be possible to
handle them for $n$ beyond 1 or 2.
Our work in second-order perturbation theory around spherical spacetimes
\cite{BMM06, BMM07} motivated us to construct such a system,
a very preliminary version of which was presented in an appendix of
\cite{BMM06}.

Here we introduce version 1.0 of the free-software package \xPert,
which implements all those expansion formulas and a number of other
efficient tools to work with metric perturbation theory in the standard
scenarios of general relativity, cosmology, quantum gravity or string
theory.
This section describes the main commands and features of the
package, simultaneously constructing a very simple example session.
(The {\em In[*]:=} and {\em Out[*]:=} prompts represent respectively
input and output lines in {\em Mathematica}. Code lines without prompt
indicate internal definitions in the package.)
Next section will provide timings with harder examples.

\xPert\/ is a module of the framework {\em xAct} \cite{xAct} for
efficient tensor computer algebra in {\em Mathematica}.
{\em xAct} is based on the use of fast algorithms of computational group
theory, which allow very fast canonicalization of indexed expressions
with arbitrary symmetries \cite{Portugal, xPerm}.
Its core tensor packages are {\em xTensor} for abstract computations
and {\em xCoba} for component computations, but {\em xAct} has also
modules for manipulation of the Riemann tensor ({\em Invar} \cite{Invar2})
or tensor spherical harmonics ({\em Harmonics} \cite{xAct}),
and others are under development.
These packages share a consistent notation, so that they can be run
together, and are all free software.

Following Section \ref{theory}, which is valid for any background metric,
\xPert\/ has been developed at the abstract computation level, to
avoid dealing with choices of coordinate systems or frames. Therefore,
it runs on {\em xTensor}, using its notations and conventions.
A tensor field $T_a{}^b$ is denoted by {\tt T[-a, b]} and the
partial derivative operator $\partial_a$ is represented as
{\tt PD[-a]}, such that $\partial_a v^b$ is
{\tt PD[-a][v[b]]} in the system. Let us start by loading
{\em xTensor}:

\mathIn{1}{<<xAct`xTensor`}
\mathComment{(Version and copyright messages)}

We first define our background structure: a four-dimensional
manifold {\tt M} whose tangent vector space will have
(abstract) indices {\tt \{a,b,c,d,e\}},

\mathIn{2}{DefManifold[ M, 4, \{a,b,c,d,e\} ]}

\noindent
Then we define a metric tensor field {\tt g} with negative determinant
and associated Levi-Civita covariant derivative {\tt CD},

\mathIn{3}{DefMetric[ -1, g[-a,-b], CD, \{";","$\nabla$"\} ]}
\mathComment{(Info messages on construction of associated tensors)}

\noindent We have provided the symbols {\tt \{";","$\nabla$"\}} to
format the derivative in postfix or prefix output notation, respectively.
{\tt DefMetric} automatically defines all tensors normally associated
to the metric or its connection, such as {\tt ChristoffelCD[a,-b,-c]},
{\tt RiemannCD[-a,-b,-c,-d]}, {\tt EinsteinCD[-a,-b]},
{\tt Detg[]}, and so on, with obvious meanings. We can define other
tensors with the syntax

\mathIn{4}{DefTensor[ MaxwellF[a,b], M, Antisymmetric[\{a,b\}], PrintAs->"F" ]}

\noindent The arrow {\tt ->} is the {\em Mathematica} representation
for an optional named argument.

Now we load \xPert\/ (this would also load automatically {\em xTensor}
if it wasn't already in memory):

\mathIn{5}{<<xAct`xPert`}
\mathComment{- - - - - - - - - - - - - - - - - - - - - - - - - - - - - - - - - - - - - - - - - - -}
\mathComment{Package xAct`xPert`\ \  version 1.0.0, \{2008, 6, 30\}}
\mathComment{Copyright (C) 2005--2008 David Brizuela, Jose M. Martin-Garcia}
\mathComment{ and Guillermo A. Mena Marugan, under GPL}
\mathComment{- - - - - - - - - - - - - - - - - - - - - - - - - - - - - - - - - - - - - - - - - - -}

\noindent This adds several new commands and reserved words to the system,
of which we shall here describe the four most important, namely
{\tt DefMetricPerturbation}, {\tt Perturbation},
{\tt ExpandPerturbation} and {\tt GaugeChange}.

A perturbative structure having metric {\tt g} as background and the
tensor {\tt h} as its perturbation is defined using

\mathIn{6}{DefMetricPerturbation[ g, h, $\varepsilon$ ]}

\noindent which also identifies {\tt $\varepsilon$} as the perturbative
parameter of the expansions. From now on, the $n$-th perturbation of the
metric {\tt g[-a,-b]} will be denoted as {\tt h[LI[n],-a,-b]},
where {\tt LI} is the {\em xTensor} head to denote so-called
`label indices', that is, indices with no vector space association.
Labels can be considered as general non-geometric purpose indices.

The $\Delta$ operator of section \ref{theory} is represented by the
head {\tt Perturbation}. It has two arguments: the background expression
being perturbed and the perturbative order (with default value 1):

\mathIn{7}{Perturbation[ MaxwellF[a,b], 3 ]}
\mathOut{7}{$\Delta^3[F^{ab}]$}

\noindent Note that the tensor is represented with its symbol $F$ and
that the perturbation order is an exponent of $\Delta$, as in section
\ref{theory}. Following normal {\em Mathematica}, the output is formatted
for the sake of clarity, but the internal notation is still the same.
{\tt Perturbation} acts mainly as a wrapper for tensor expressions, but
has been instructed to evaluate them under certain conditions. First,
it automatically combines perturbative orders of composed heads  (symbols
with an underscore are named patterns in {\em Mathematica}):

\mathCode{Perturbation[ expr\_, 0 ] := expr}
\mathCode{Perturbation[ Perturbation[ expr\_, n\_ ], m\_ ] :=
          Perturbation[ expr, n+m ]}

\noindent Being a derivative, {\tt Perturbation} is linear and gives zero
on the {\tt delta} identity tensor and constants:

\mathCode{Perturbation[ x\_ + y\_, n\_ ] := Perturbation[ x, n ] +
          Perturbation[ y, n ]}
\mathCode{Perturbation[ delta[a\_, b\_], n\_ ] := 0}
\mathCode{Perturbation[ expr\_?ConstantQ, n\_ ] := 0}

\noindent The Leibnitz rule is also automatic, and has been implemented
following equation (\ref{product}) for any number of factors and any
perturbative order, using fast algorithms to compute partitions
implemented in \xPert. {\tt Perturbation} commutes with
partial derivatives of general expressions and with any covariant
derivative of a scalar expression:

\mathCode{Perturbation[ PD[-a\_][ expr\_ ], n\_ ] :=
          PD[-a][ Perturbation[ expr, n ] ]}
\mathCode{Perturbation[ CD\_?CovDQ[-a\_][ expr\_?ScalarQ ], n\_ ] :=}
\mathCode{\qquad  CD[-a][ Perturbation[ expr, n ] ]}

\noindent The index of the derivatives is required to be always covariant,
to avoid a metric mismatch, and that is implemented through a pattern
index {\tt -a\_} . Finally, {\tt Perturbation} does not change the density
weight of the perturbed expression:

\mathCode{WeightOf[ Perturbation[ expr\_, n\_ ] ] := WeightOf[ expr ]}

The {\tt DefMetricPerturbation} in {\em In[6]} defines special rules
for the metric {\tt g} and its perturbations {\tt h} with covariant
indices:

\mathCode{Perturbation[ g[-a\_,-b\_], n\_ ] := h[LI[n],-a,-b]}
\mathCode{Perturbation[ h[LI[n\_],-a\_,-b\_], m\_ ] :=
          h[LI[n+m],-a,-b]}

\noindent With the setup and internal definitions so far we can now
perform computations like this second order perturbation

\mathIn{8}{Perturbation[ g[-a,-b] RicciCD[c,d] +
             RiemannCD[-a,-b,c,d], 2 ]}
\mathOut{8}{$2\, h^1{}_{ab}\, \Delta[R_{cd}] + g_{ab}\, \Delta^2[R_{cd}]
           +\Delta^2[R_{ab}{}^{cd}] + h^2{}_{ab}\,R_{cd}$}

\noindent Actually, we could now proceed to perform any computation
in metric perturbation theory by decomposing the curvature tensors
in partial derivatives of the metric and using the code already
given recursively. Only the definition
$\Delta[g^{ab}] = - \pert{1}{h}^{ab}$ would be missing.
However, that would be highly inefficient already for moderate perturbative
order $n$. Instead, we shall use the expansion formulas of section
\ref{theory}, which allow the nonrecursive construction of perturbations
at any order $n$.

Formulas (\ref{pertCD}, \ref{pertLieD}) for derivative expansions
and formulas (\ref{invmetricexpansion}, \ref{Christoffelexpansion},
\ref{pertRiemann2}, \ref{pertRicci}, \ref{pertRicciScalar},
\ref{pertEinstein}, \ref{pertDet})
for the relevant curvature tensors have all been encoded in a single
command called {\tt ExpandPerturbation}, the most powerful part of
\xPert. {\tt ExpandPerturbation} takes any expression and replaces
the arbitrary-order perturbations of known background objects by their
expansions in terms of metric perturbations, but only if those objects
have their indices in the appropriate positions. For example the
perturbation of the Einstein tensor has only been stored for covariant
indices. In all other cases there is an internal call to the {\em xTensor}
function {\tt SeparateMetric}, which introduces metric factors to bring
all indices to their {\em natural} positions, which are those given
at definition time. To show how this works, we perform an explicit
metric separation by hand (symbol {\tt\%} represents the previous
output): 

\mathIn{9}{Perturbation[ EinsteinCD[a,b] ]}
\mathOut{9}{$\Delta[G^{ab}]$}

\mathIn{10}{SeparateMetric[ ][ \% ]}
\mathOut{10}{$G_{cd}\,g^{bd}\,\Delta[g^{ac}] +
    g^{ac}\left(\,g^{bd}\,\Delta[G_{cd}]+G_{cd}\,\Delta[g^{bd}]\,\right)$}

\noindent Now {\tt ExpandPerturbation} can expand the perturbation of
the Einstein tensor with covariant indices, and the perturbation of the
inverse metric.

\mathIn{11}{ContractMetric[ ExpandPerturbation[ \% ] ]}
\mathOut{11}{$-G_c{}^bh^{1ac}
            -G^a{}_ch^{1bc}
            +\frac{1}{2}g^{ab}h^{1cd}R_{cd}
            -\frac{1}{2}h^{1ab}R
            -\frac{1}{2}h^{1c}{}_c{}^{;b;a}$}
\mathComment{$-\frac{1}{2}h^{1cb}{}_{;c}{}^{;a}
            +\frac{1}{2}h^{1b}{}_c{}^{;c;a}
            +\frac{1}{2}h^{1cb;a}{}_{;c}
            +\frac{1}{2}h^{1ca;b}{}_{;c}
            -\frac{1}{2}h^{1ba;c}{}_{;c}$}
\mathComment{$+\frac{1}{4}g^{ab}h^{1d}{}_c{}^{;c}{}_{;c}
            +\frac{1}{4}g^{ab}h^{1dc}{}_{;d;c}
            -\frac{1}{4}g^{ab}h^{1c}{}_c{}^{;d}{}_{;c}
            -\frac{1}{2}g^{ab}h^{1dc}{}_{;c;d}
            +\frac{1}{4}g^{ab}h^{1c}{}_c{}^{;d}{}_{;d}$}

\noindent The {\em xTensor} command {\tt ContractMetric} has been used
to absorb all possible metric factors. Finally, {\tt ToCanonical}
moves indices around bringing equal terms together,

\mathIn{12}{ToCanonical[ \% ]}
\mathOut{12}{$-G^{bc}h^{1a}{}_c
            -G^{ac}h^{1b}{}_c
            +\frac{1}{2}g^{ab}h^{1cd}R_{cd}
            -\frac{1}{2}h^{1ab}R
            -\frac{1}{2}h^{1c}{}_c{}^{;a;b}$}
\mathComment{$+\frac{1}{2}h^{1bc;a}{}_{;c}
            +\frac{1}{2}h^{1ac;b}{}_{;c}
            -\frac{1}{2}h^{1ab;c}{}_{;c}
            -\frac{1}{2}g^{ab}h^{1cd}{}_{;c;d}
            +\frac{1}{2}g^{ab}h^{1c}{}_c{}^{;d}{}_{;d}$}

\begin{figure}
\includegraphics[width=17cm]{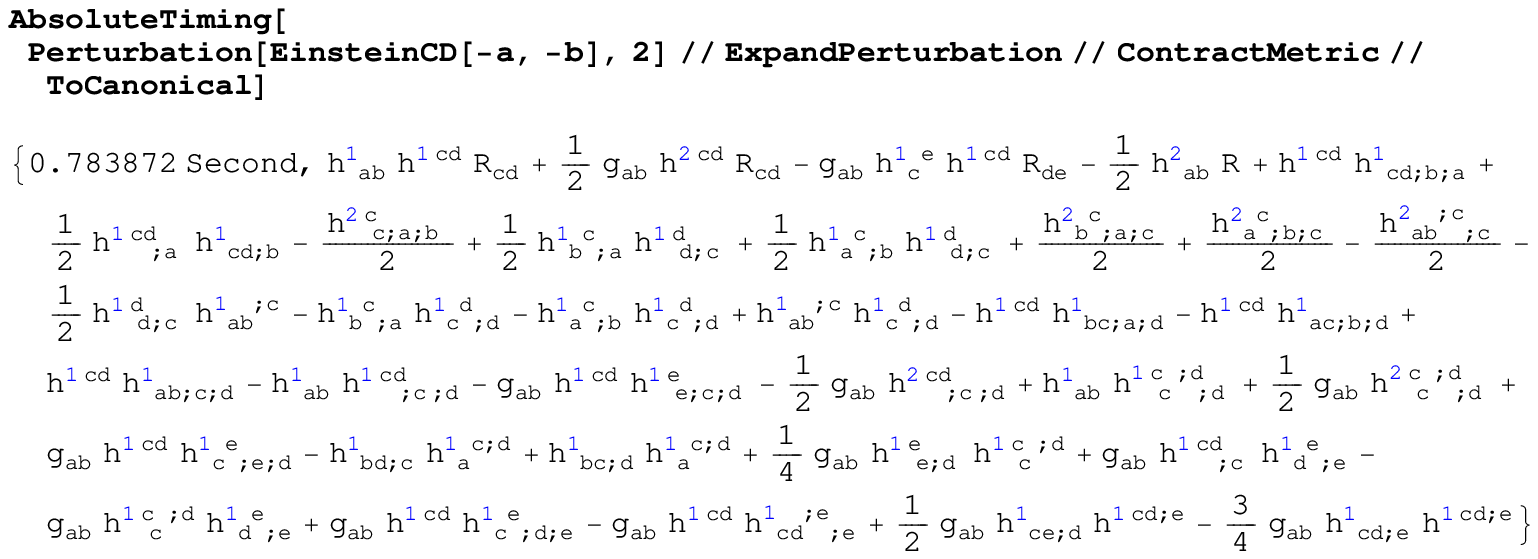}
\caption{\label{einstein}
The second-order perturbation of the Einstein tensor is constructed and
canonicalized in less than one second. The blue labels of the
{\tt h} tensors denote the perturbative order.}
\end{figure}

\noindent Figure \ref{einstein} shows the output of the second perturbation
of the covariant Einstein tensor, computed with the same combination of
commands.

Another useful command implemented in \xPert\/ is {\tt GaugeChange}.
A $\varepsilon$-dependent family of diffeomorphisms on a manifold
can be parametrized by an infinite collection of vector fields
$\pert{n}{\xi}^a$, one per perturbative order, and the action of
these on any background tensor at any order has been given by Bruni
and collaborators \cite{BMM97}. The general case has been implemented
here and can be used as follows. First define the family of generator
vector fields on the manifold {\tt M} transforming from the current
gauge to a new gauge:

\mathIn{13}{DefTensor[ $\xi$[LI[n],a], M ]}

\noindent The third-order perturbation of $F^{ab}$ can be changed to
the new gauge using

\mathIn{14}{GaugeChange[ Perturbation[ MaxwellF[a,b], 3 ], $\xi$ ]}
\mathOut{14}{$\Delta^3[F^{ab}] 
+3\,{\cal L}_{\xi^1}\Delta^2[F^{ab}]
+3\,{\cal L}_{\xi^1}{\cal L}_{\xi^1}\Delta[F^{ab}]
+ {\cal L}_{\xi^1}{\cal L}_{\xi^1}{\cal L}_{\xi^1} F^{ab}$}
\mathComment{$
+3\,{\cal L}_{\xi^1}{\cal L}_{\xi^2}F^{ab}
+3\,{\cal L}_{\xi^2}\Delta[F^{ab}]
+{\cal L}_{\xi^3}F^{ab}$}


We finish this section by coming back to the problem of perturbation
theory using the background field method, in which all but the first metric
perturbations vanish, as stated in section \ref{theory}. This can be
easily implemented setting

\mathIn{15}{h[LI[n\_],a\_,b\_]:=0 /; n > 1}

\section{Examples}\label{examples}

This section presents several examples of computations, focusing on the
dependence of their timings on the perturbative order and the number
of objects being perturbed. The intrinsic combinatorial nature of the
problem will always imply exponential dependence, but we will see that
the timings in \xPert\/ are short enough to handle all useful cases.
These examples have been performed using a Linux box with a 3 GHz
Pentium IV processor and 2 Gb of RAM.

\begin{figure}
\includegraphics[width=10cm]{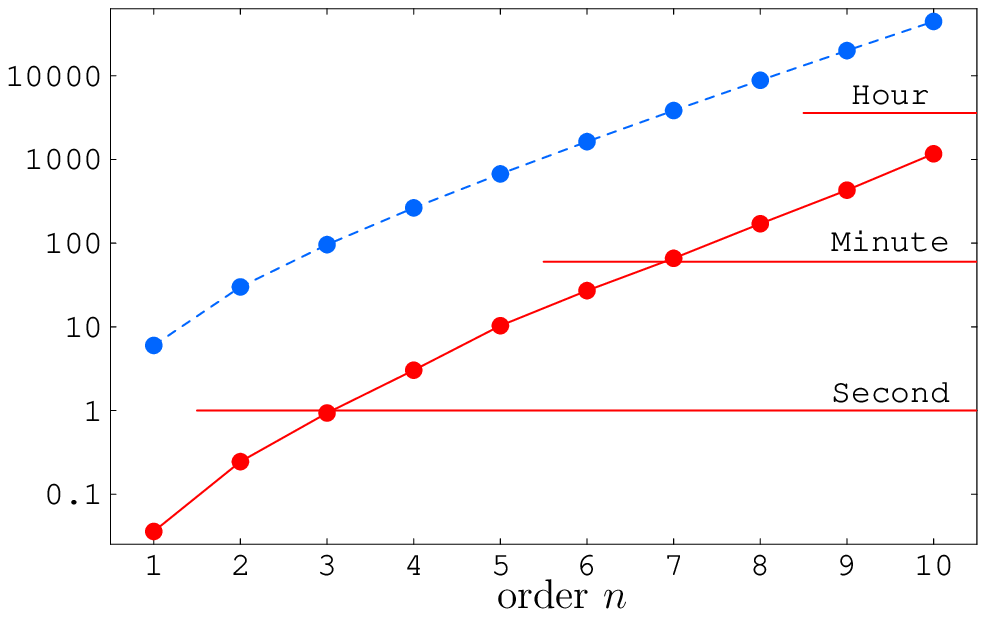}
\caption{\label{riemann}
Canonicalization timings (in seconds) for the perturbation of the
Riemann tensor at perturbative orders $n=1...10$ (lower, red line).
Also shown number of terms in the expression (upper, blue dashed line).
Both lines are clear exponentials in $n$, with the timings growing
slightly faster because terms with larger $n$ are harder to canonicalize
due to their larger average number of indices.
}
\end{figure}

In perturbation theory the overall level of complexity is mainly
determined by the perturbative order $n$. It affects the computation
in two different ways: on the one hand the expressions to manipulate
are sums with a number of terms which grows exponentially with $n$;
on the other hand each term is a product of objects and the
number of factors also grows (typically linearly) with $n$.
Canonicalizing a sum of terms is obviously a linear process because
each term can be dealt with independently (this is an ideal scenario
for parallelization), but canonicalization of a product of objects
is naturally factorial in the number of indices and this could
prevent any practical computation. The canonicalization algorithms in
{\em xTensor} are fast enough to render the problem effectively
polynomial in the number of indices, allowing us to deal with expressions
of a few dozen indices in hundredths of a second.
Figure \ref{riemann} shows the number
of terms and the timing of canonicalization of the perturbation of the
Riemann tensor at different perturbative orders. (The timing of
construction of the expression is negligible in comparison.)
The 10-th order perturbation is canonicalized in slightly less than
20 minutes and contains 44544 terms. We see clearly the exponential growth
of both curves, but with manageable timings. Third-order perturbation
expressions can be manipulated in 1 second.

\begin{figure}
\includegraphics[width=10cm]{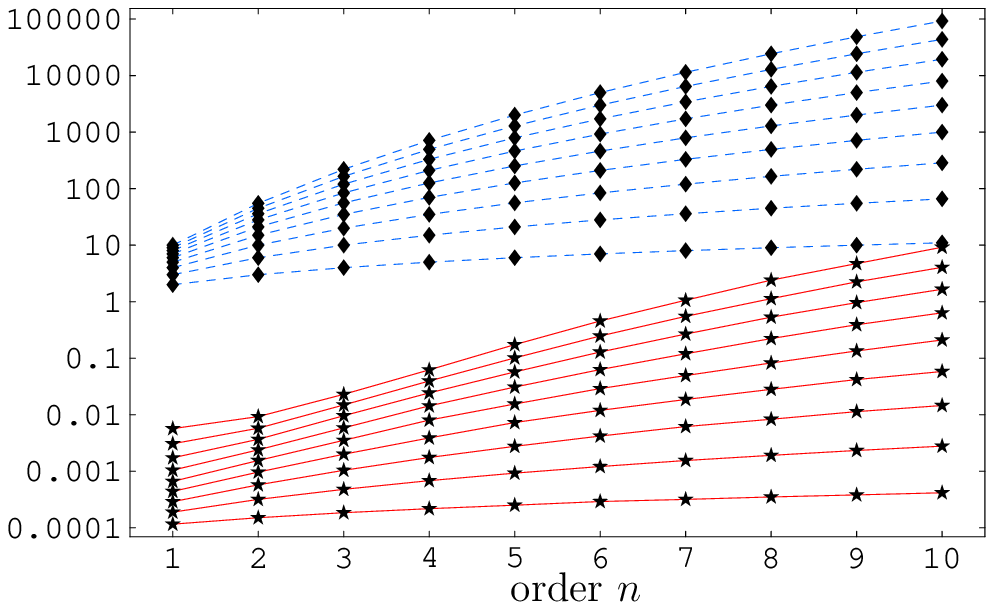}
\caption{\label{leibnitz}
Timings of expansion (in seconds; red lines, stars) and number of terms
(blue dashed lines, diamonds) of the perturbation of the product of $m$
factors, for different perturbative orders $n=1...10$ [formula
(\ref{product})]. Different lines correspond to increasing values of $m$,
from $m=2$ to $m=10$ starting from below. All practical cases stay below
one second.
}
\end{figure}

\begin{figure}
\includegraphics[width=10cm]{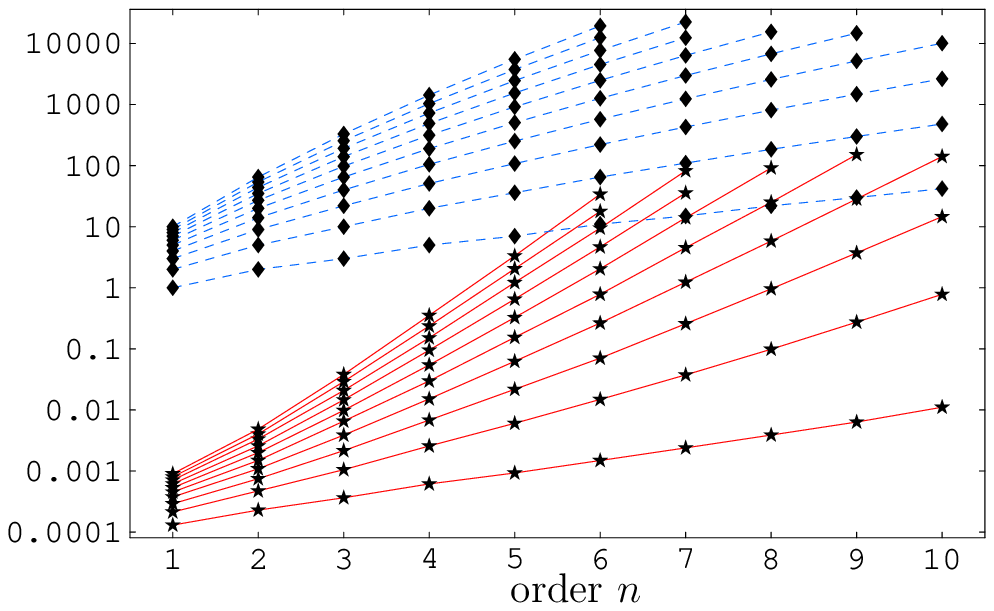}
\caption{\label{chain}
Timings of expansion (in seconds; red lines, stars) and number of terms
(blue dashed lines, diamonds) of the perturbation of a scalar function of
$m$ scalar arguments, for different perturbative orders $n=1...10$.
Different lines correspond to increasing values of $m$, from $m=1$ to
$m=10$ starting from below. We include only those cases which can be
handled with 2Gb of RAM memory, corresponding to a few tens thousand
terms. For example, for $n=10$ we can only handle up to $m=4$.
}
\end{figure}

Other possible sources of complexity in perturbative computations are
the expansion of perturbations of a product of tensors and
the expansion of perturbations of a function of a number of scalar arguments.
Our implementations of the $n$th-order Leibnitz rule and Fa\`a di Bruno
formula, respectively, are fast enough to neglect their timings
in comparison with those of canonicalization. Figures \ref{leibnitz}
and \ref{chain} display example timings for those problems. The Leibnitz
rule is simpler than the Fa\'a di Bruno formula and produces faster results,
also taking less memory.

Overall, we see that we are limited in size by RAM memory, which allows us
to work with up to $10^5$ terms roughly with a few Gbytes, corresponding
to $n=10$ approximately. Time limitations come mainly from the
canonicalization process (other expansions are faster): Within seconds we
can manipulate all equations up to orders $n=4$ or $n=5$. The $n=10$
equations require canonicalization times of the order of 1 hour. This
gives an idea of the power and efficiency of \xPert, and what can
be achieved with it.

\section{Conclusions}
\label{conclusions}

Linear perturbation theory is an important tool in general relativity,
nowadays playing a complementary role to numerical relativity \cite{Seidel}.
Its precision and applicability can be extended by adding higher-order
perturbations in the computation, transforming the intrinsic non-linearity
of the theory into sources for the linearized equations of motion.
In most cases these sources are huge, and hence practical calculations
require the use of specialized computer algebra to construct and
manipulate them.

In this article we have introduced version 1.0 of the package
\xPert\/ for high-order metric perturbation theory. This is a very hard
problem, of intrinsic combinatorial nature, and hence exponential in
the perturbative order $n$. We have developed and coded up highly
efficient algorithms and tools, based on the combination of
explicit non-recursive formulas to build the equations at any order
$n$, and powerful algorithms of computational group theory to manipulate
them. \xPert\/ can work both with a perturbative scheme in which the
metric is expanded as an infinite series of perturbations (the usual
approach in general relativity) and with the alternative scheme in
which the metric is written as a background plus a single `deviation'
term (as used in quantum field theory).
\xPert\/ allows working with orders $n=4$ and $n=5$ in seconds,
and we have shown that it is possible to reach $n=10$ with average timings
below 1 hour with moderate computer resources. The main limitation (both
in time and memory) is the huge number of terms produced, but this can
be easily dealt with using parallelization, because different terms in
the expression can be treated independently by different processors.

\xPert\/ is a module of the general framework {\em xAct} \cite{xAct}
for tensor computer algebra in {\em Mathematica}, sharing a consistent
notation with the whole system. This means that once the perturbative
equations have been constructed, there are several hundred additional
commands available for tensor manipulations, including particularization
to a special background metric, and several thousand more for generic
manipulation in {\em Mathematica}. In particular, if one wants to
convert the produced equations into C or FORTRAN code, it can be easily
achieved with the {\tt CForm} and {\tt FortranForm} commands in
{\em Mathematica}.

\xPert\/ has been already used in several research projects: 
Our construction of a general formalism for high-order metric perturbations
around spherical backgrounds \cite{BMM06} and its gauge-invariant form
\cite{BMM07}. In a cosmological setting, high-order perturbations of
non-linear radiation transfer have been studied using \xPert, both in
kinetic theory \cite{Pit07} and in a fluid approximation \cite{Pit08},
as well as first-order perturbations of scalar field inflation for
anisotropic spacetimes \cite{CPU08}. 


The package \xPert\/ is free software distributed under the GNU
general public license, and can be download from the webpage
{\tt http://metric.iem.csic.es/Martin-Garcia/xAct/xPert/} .

\acknowledgments

We thank Cyril Pitrou for discussions and suggestions to
improve \xPert.
D.B. acknowledges financial support from the FPI
program of the Regional Government of Madrid.
J.M.M.-G. acknowledges financial aid provided by the
I3P framework of CSIC and the European Social Fund.
He also whishes to thank Vetenskapsr\aa det (Swedish Research
Council) for supporting a visit to Link\"opings universitet,
and the Mathematics Department for their hospitality.
This work was supported by the Spanish MEC Project
FIS2005-05736-C03-02 (with its continuation FIS2008-06078-C03-03).

\end{document}